\documentstyle[prl,aps]{revtex}
\begin{document}
\draft
\preprint{Dortmund, January 1994}

%  for the \twocolum option
%\advance\hsize0.5truecm\advance\hoffset-0.25truecm
%\advance\columnsep-0.5truecm\twocolumn[

\title{Equation of motion approach to the Hubbard model
in infinite dimensions}
\author{Claudius Gros\thanks{e-mail: UPH301 at DDOHRZ11},
       }
\address{Institut f\"ur Physik, Universit\"at Dortmund,
         44221 Dortmund, Germany
        }
\date{\today}
\maketitle
\begin{abstract}

%  for the \twocolum option
%\hbox to \hsize{\hfill\vbox{\hsize=14truecm

We consider the Hubbard model on the
infinite-dimensional Bethe lattice and
construct a systematic series of
self-consistent approximations
to the one-particle Green's function,
$G^{(n)}(\omega),\ n=2,3,\dots\ $ .
The first $n-1$ equations of motion
are exactly fullfilled by
$G^{(n)}(\omega)$ and the
$n$'th equation of motion is decoupled
following a simple set of decoupling rules.
$G^{(2)}(\omega)$ corresponds to the
Hubbard-III approximation. We present
analytic and numerical results for the
Mott-Hubbard transition at half filling
for $n=2,3,4$.

%  for the \twocolum option
%             }\hfill}
\end{abstract}
%%%%%%%%%%%%%%%%%%%%%%%%%%%%%%%%%%%%%%%%%%%
%%%% end bracket of \twocolumne option %%%%
%%%%%%%%%%%%%%%%%%%%%%%%%%%%%%%%%%%%%%%%%%%
%\maketitle
%]

%
%%%%%%%%%%%%%%%%%%%%%%%%%%%%%%%%%%%%%%%%%%%%%%%%%%%%%%%
%
%%%%%%%%%%%%%%%%%%%%%%%%%%%%%%%%%%%%%%%%%%%%%%%%%%%%%%%
%
%\narrowtext
\advance\baselineskip by 5pt
\renewcommand{\arraystretch}{2.0} \jot=8pt

%%%%% the * suppresses automatic numbering
\section*{Introduction}

The Hubbard-model on the infinite-dimensional,
half-filled Bethe lattice in the paramagnetic state
has been considered in the last years
more and more as
the standard model for the Mott-Hubbard transition.
One of the reasons for this interest is that
the study of interacting Fermions in the
limit of infinite spatial dimensions
\cite{Vollhardt_review} leads to
considerable technical advantages, as the
self-energy becomes stricly local. The
Mott-Hubbard transition has been studied
recently by Monte-Carlo \cite{Monte_Carlo},
a self-consistent weak-coupling theory \cite{IPT},
rigorous \cite{star_of_stars} and self-consistent
exact diagonalization studies \cite{exact_dia},
with considerable different results,
see \cite{star_of_stars}.

Alternatively, Hubbard \cite{Hubbard_III} has
considered an equation of motion approach to the
Hubbard model. Unfortunatly his approach can not
be systematically improved in any finite dimensions,
since the resulting self-consitency equations would be
numerically intracable, due to the involved
summations over momenta. Here we point out
that, uniquely to the infinite-dimensional
Bethe lattice, the equation of motion approach
may be systematically carried on. We develop
a simple decoupling scheme applicable
to equations of motion in any orders and
resulting in a self-consistency equation for
the one-particle Green's function.
We present results for the Green's
function is second, third and fourth order,
with the second order corresponding to the
Hubbard-III solution \cite{Hubbard_III}.
%
%%%%%%%%%%%%%%%%%%%%%%%%%%%%%%%%%%%%%%%%%%%%%%%%%%%%%%%
%
%%%%%%%%%%%%%%%%%%%%%%%%%%%%%%%%%%%%%%%%%%%%%%%%%%%%%%%
%

\section*{Definitions}

At half filling, $ n=1$, the chemical potential
$\mu\equiv U/2$. For general fillings we set
$\mu=U/2+\Delta\mu$ and find for the grand-canonical
Hubbard Hamiltonian,
$\hat K = \hat H - \mu\hat N$,
\begin{equation}
\hat K\ = \ t\sum_{<i,j>,\sigma}
          \hat t_{j,i,\sigma}^{\phantom{\dagger}}
      \ +\ U/2\sum_i
          (\hat u_i^{\phantom{\dagger}} -1)
      \ -\ \Delta\mu\sum_i
          \hat n_i^{\phantom{\dagger}},
\label{K}
\end{equation}
where the symbol $<i,j>$ denotes
pairs of nearest neighbour sites on a Bethe-lattice
with coordination number $z$. The scaling
$t=\tilde t/\sqrt z$ yields a non-trivial \cite{Metzner_Vollhardt}
limiting behaviour in the limit $z\rightarrow\infty$.
In Eq.\ (\ref{K}) we have made use of
$\hat n_{i  ,\sigma}^{\phantom{\dagger}} =
 \hat c_{i,\sigma}^{\dagger}\hat c_{i,\sigma}^{\phantom{\dagger}},\
\hat n_{i         }^{\phantom{\dagger}}  =
 \hat n_{i,\uparrow}^{\phantom{\dagger}}
+\hat n_{i,\downarrow}^{\phantom{\dagger}}$
in Eq.\ (\ref{K}) and of some of
the following operator definitions:
\begin{equation}
\begin{array}{rcl}
\hat t_{j,i,\sigma}^{\phantom{\dagger}}
                   \ &=& \
 \hat c_{j,\sigma}^{\dagger}\hat c_{i,\sigma}^{\phantom{\dagger}}
+\hat c_{i,\sigma}^{\dagger}\hat c_{j,\sigma}^{\phantom{\dagger}} \\
 \hat j_{j,i,\sigma}^{\phantom{\dagger}}
                   \ &=& \
 \hat c_{j,\sigma}^{\dagger}\hat c_{i,\sigma}^{\phantom{\dagger}}
-\hat c_{i,\sigma}^{\dagger}\hat c_{j,\sigma}^{\phantom{\dagger}} \\
 \hat u_{i       }^{\phantom{\dagger}}
                   \ &=& \
 \hat c_{i,\uparrow}^{\dagger}
 \hat c_{i,\uparrow}^{\phantom{\dagger}}
 \hat c_{i,\downarrow}^{\dagger}
 \hat c_{i,\downarrow}^{\phantom{\dagger}}
+\hat c_{i,\uparrow}^{\phantom{\dagger}}
 \hat c_{i,\uparrow}^{\dagger}
 \hat c_{i,\downarrow}^{\phantom{\dagger}}
 \hat c_{i,\downarrow}^{\dagger} \\
\hat d_{i  ,\sigma}^{\phantom{\dagger}}
                   \ &=& \
 \hat c_{i,\sigma}^{\dagger}\hat c_{i,\sigma}^{\phantom{\dagger}}
-\hat c_{i,\sigma}^{\phantom{\dagger}}\hat c_{i,\sigma}^{\dagger}
= 2\hat n_{i  ,\sigma}^{\phantom{\dagger}} -1,
\label{definitions}
\end{array}
\end{equation}
where the $\hat c_{i,\sigma}^{\dagger}$
and the $\hat c_{i,\sigma}^{\phantom{\dagger}}$
are creation/destruction operators for electrons with spin
$\sigma=\uparrow,\downarrow$
on lattice sites $i$. The following operator identities hold:
\begin{eqnarray}
\hat c_{j,\sigma}^{\dagger}
\hat c_{i  ,\sigma}^{\phantom{\dagger}} \ &=& \
(\hat t_{j,i,\sigma}^{\phantom{\dagger}}
+\hat j_{j,i,\sigma}^{\phantom{\dagger}} )/2
         \label{identity_1} \\
 \hat d_{i,\sigma}^{\phantom{\dagger}}
 \hat d_{i,\sigma}^{\phantom{\dagger}} \ &\equiv &\ \hat 1.
 \label{identity_2}
\end{eqnarray}
For the equation of motion we will make use of
following operator commutator relations:
\begin{eqnarray}
\left[
 \hat t_{j,i,\sigma}^{\phantom{\dagger}}
,\hat c_{i,\sigma}^{\phantom{\dagger}} \right] \ &=& \
-\hat c_{j,\sigma}^{\phantom{\dagger}}
 \label{commutator_1} \\
\left[
 \hat u_{i       }^{\phantom{\dagger}}
,\hat c_{i,\sigma}^{\phantom{\dagger}} \right] \ &=& \
-\hat d_{i,-\sigma}^{\phantom{\dagger}}
 \hat c_{i,\sigma}^{\phantom{\dagger}}
 \label{commutator_2} \\
\left[
 \hat t_{j,i,-\sigma}^{\phantom{\dagger}}
,\hat d_{i,-\sigma}^{\phantom{\dagger}} \right] \ &=& \
2\,\hat j_{j,i,-\sigma}^{\phantom{\dagger}}
 \label{commutator_3} \\
\left[
 \hat u_{j       }^{\phantom{\dagger}}
,\hat j_{j,i,-\sigma}^{\phantom{\dagger}} \right] \ &=& \
 \hat d_{j,\sigma}^{\phantom{\dagger}}
 \hat t_{j,i,-\sigma}^{\phantom{\dagger}}
 \label{commutator_4} \\
\left[
 \hat t_{j,i,-\sigma}^{\phantom{\dagger}}
,\hat j_{j,i,-\sigma}^{\phantom{\dagger}} \right] \ &=& \
 2(\hat n_{i,-\sigma}^{\phantom{\dagger}}
 - \hat n_{j,-\sigma}^{\phantom{\dagger}}),
 \label{commutator_5} \\
\left[
 \hat u_{i}^{\phantom{\dagger}}
,\hat j_{j,i,-\sigma}^{\phantom{\dagger}}
 \hat c_{i,\sigma}^{\phantom{\dagger}} \right] \ &=& \ 0 \ =\
\left[
 \hat u_{i}^{\phantom{\dagger}}
,\hat t_{j,i,-\sigma}^{\phantom{\dagger}}
 \hat c_{i,\sigma}^{\phantom{\dagger}} \right],
 \label{commutator_6} \\
\end{eqnarray}
and similar one's.
Given operators $\hat A$ and $\hat B$, the Matsubara Green's function
$\langle\langle\hat A;\hat B\rangle\rangle$ is defined as
\[
\,\langle\langle\hat A;\hat B\rangle\rangle\ =\ \int_o^{\beta}d\tau
  (-1)<T_{\tau}\hat A(\tau)\hat B(0)>
      e^{i\omega_n\tau},
\]
with $\tau$ being the imaginary time and the brackets
$<\dots>$ denoting the thermodynamic expectation value.
The retarded Green's function
is obtained via the analytic continuation
$i\omega_n\rightarrow\omega+i\delta$.
Here we are interested in constructing systematic approximations to
the onsite one-particle Green's function
$G(i\omega_n)=\,\langle\langle\hat c_{0,\sigma}^{\phantom{\dagger}};
  \hat c_{0,\sigma}^{\dagger}\rangle\rangle$.
In the following we will
define sites $1,\ 1^{\prime},\ 1^{\prime\prime},\dots$ to be
different n.n. sites of the central site, 0, and
$2,\ 2^{\prime},\ 2^{\prime\prime},\dots$ to be
different n.n.n. sites of the central site.

%
%%%%%%%%%%%%%%%%%%%%%%%%%%%%%%%%%%%%%%%%%%%%%%%%%%%%%%%
%
%%%%%%%%%%%%%%%%%%%%%%%%%%%%%%%%%%%%%%%%%%%%%%%%%%%%%%%
%

\section*{Equations of motion}

For any imaginary-time-dependent operator, $\hat A(\tau)$ in
the Heisenberg picture, the equation of motion is given by
\[
{d\over d\tau}\hat A(\tau) \ = \ [\hat K, \hat A],
\]
where, in the grand-canonical formulation, $\hat K=\hat H-\mu\hat N$
is given by Eq.\ (\ref{K}). Note that we have set
$\mu=U/2+\Delta\mu$.
Some examples of specific equation of motion in
$\tau$-space are
\begin{equation}
\begin{array}{rcl}
{d\over d\tau} \hat c_{i,\sigma}^{\dagger} (\tau)\ &=& \
  -\Delta\mu\, \hat c_{i,\sigma}^{\dagger} (\tau)
         +  t\sum_j \hat c_{j,\sigma}^{\dagger} (\tau)
                              + \
\frac{\displaystyle U}{\displaystyle 2}
                 \hat d_{i,-\sigma}^{\phantom{\dagger}}
                 \hat c_{i,\sigma}^{\dagger} (\tau)    \\
{d\over d\tau}   \hat c_{i,\sigma}^{\phantom{\dagger}} (\tau)\ &=& \
\phantom{-}
    \Delta\mu\,  \hat c_{i,\sigma}^{\phantom{\dagger}} (\tau)
        -t\sum_j \hat c_{j,\sigma}^{\phantom{\dagger}} (\tau)
                              - \
\frac{\displaystyle U}{\displaystyle 2}
                 \hat d_{i,-\sigma}^{\phantom{\dagger}}
                 \hat c_{i,\sigma}^{\phantom{\dagger}} (\tau) \\
{d\over d\tau} \hat d_{i,-\sigma}
               \hat c_{i,\sigma}^{\phantom{\dagger}} (\tau)\ &=& \
\phantom{-}
    \Delta\mu\,\hat d_{i,-\sigma}
               \hat c_{i,\sigma}^{\phantom{\dagger}} (\tau)
         +       t \sum_j\left[
                2\,\hat j_{j,i,-\sigma}^{\phantom{\dagger}}
               \hat c_{i,\sigma}^{\phantom{\dagger}} (\tau)
                -\hat d_{i,-\sigma}^{\phantom{\dagger}}
               \hat c_{j,\uparrow}^{\phantom{\dagger}} (\tau)
                         \right]
                -
\frac{\displaystyle U}{\displaystyle 2}
               \hat c_{i,\sigma}^{\phantom{\dagger}} (\tau)
\label{motion_0}
\end{array}
\end{equation}
where $j$ is a n.n. site of $i$ and where we have used
definitions Eq.\ (\ref{definitions}) and the operator identity
Eq.\ (\ref{identity_2}).
In Fourier-space the first equation of motion is
\[
(i\omega_n\ +\Delta\mu)
\,\langle\langle\hat c_{0,\sigma}^{\phantom{\dagger}};
                \hat c_{0,\sigma}^{\dagger}\rangle\rangle
     \ =\ 1 + z t
\,\langle\langle\hat c_{1,\sigma}^{\phantom{\dagger}};
                \hat c_{0,\sigma}^{\dagger}\rangle\rangle
         + U/2
\,\langle\langle\hat d_{0,-\sigma}^{\phantom{\dagger}}
 \hat c_{0,\sigma}^{\phantom{\dagger}};
                \hat c_{0,\sigma}^{\dagger}\rangle\rangle,
\]
where the site 1 is any of the $z$ equivalent
n.n. site of the central site 0. In the
limit $z\rightarrow\infty$ the decoupling
\begin{equation}
\,\langle\langle\hat c_{1,\sigma}^{\phantom{\dagger}};
                \hat c_{0,\sigma}^{\dagger}\rangle\rangle
\ \rightarrow\  t
\,\langle\langle\hat c_{1,\sigma}^{\phantom{\dagger}};
                \hat c_{1,\sigma}^{\dagger}\rangle\rangle
\,\langle\langle\hat c_{0,\sigma}^{\phantom{\dagger}};
                \hat c_{0,\sigma}^{\dagger}\rangle\rangle
\label{d_1}
\end{equation}
is exact since all diagrams contributing to the propagation of
the particle from site 1 to site 1 have only a vanishing
probability $1/z\rightarrow 0$ to visit the central site.
Using that
$
\,\langle\langle\hat c_{1,\sigma}^{\phantom{\dagger}};
                \hat c_{1,\sigma}^{\dagger}\rangle\rangle
=
\,\langle\langle\hat c_{0,\sigma}^{\phantom{\dagger}};
                \hat c_{0,\sigma}^{\dagger}\rangle\rangle
\equiv G(i\omega_n)
$
in the paramagnetic state and that $zt^2=\tilde t^2$ we
rewrite the equation of motion of first and second order
as
\begin{eqnarray}
\left[i\omega_n+\Delta\mu-\tilde t^2 G(i\omega_n)\right]
                           G(i\omega_n)  \ &=&\ 1 + U/2
\,\langle\langle\hat d_{0,-\sigma}^{\phantom{\dagger}}
 \hat c_{0,\sigma}^{\phantom{\dagger}};
                \hat c_{0,\sigma}^{\dagger}\rangle\rangle
\label{motion_1}  \\
(i\omega_n+\Delta\mu)\,\langle\langle
                \hat d_{0,-\sigma}^{\phantom{\dagger}}
 \hat c_{0,\sigma}^{\phantom{\dagger}};
                \hat c_{0,\sigma}^{\dagger}\rangle\rangle
                                         \ &=&\
 2\,\Delta n_{0,-\sigma}
  \,+\,U/2\, G(i\omega_n)
\label{motion_2}  \\                     &-&\ 2\, z t
\,\langle\langle\hat j_{1,0,-\sigma}^{\phantom{\dagger}}
 \hat c_{0,\sigma}^{\phantom{\dagger}};
                \hat c_{0,\sigma}^{\dagger}\rangle\rangle
               +\, z t
\,\langle\langle\hat d_{0,-\sigma}^{\phantom{\dagger}}
 \hat c_{1,\sigma}^{\phantom{\dagger}};
                \hat c_{0,\sigma}^{\dagger}\rangle\rangle,
\nonumber
\end{eqnarray}
where
$<\hat c_{0,-\sigma}^{\dagger}
  \hat c_{0,-\sigma}^{\phantom{\dagger}}
 -\hat c_{0,-\sigma}^{\phantom{\dagger}}
  \hat c_{0,\sigma}^{\dagger}>\,
 = 2\,n_{0,-\sigma}-1\equiv 2\Delta n_{0,-\sigma}$
and where we have measured $n_{0,-\sigma}\equiv
1/2+\Delta n_{0,-\sigma}$ with respect to half-filling.
The second-order equation of motion, Eq.\ (\ref{motion_2}),
generates two new Green's function. In third order
we have the equation of motion for
$\,\langle\langle\hat j_{1,0,-\sigma}^{\phantom{\dagger}}
 \hat c_{0,\sigma}^{\phantom{\dagger}};
                 \hat c_{0,\sigma}^{\dagger}\rangle\rangle$,
\begin{equation}
\begin{array}{rcl}
(i\omega_n+\Delta\mu)\,\langle\langle
              \hat j_{1,0,-\sigma}^{\phantom{\dagger}}
 \hat c_{0,\sigma}^{\phantom{\dagger}};
              \hat c_{0,\sigma}^{\dagger}\rangle\rangle
     \ &=&\
 \,\langle\langle\hat c_{1,-\sigma}^{\dagger}
              \hat c_{0,-\sigma}^{\phantom{\dagger}}
 -\hat c_{0,-\sigma}^{\dagger}
              \hat c_{1,-\sigma}^{\phantom{\dagger}}\rangle\rangle
        +\,z t
\,\langle\langle\hat j_{1,0,-\sigma}^{\phantom{\dagger}}
 \hat c_{1^{\prime},\sigma}^{\phantom{\dagger}};
              \hat c_{0,\sigma}^{\dagger}\rangle\rangle
 \\  \ &-&\ z t
\,\langle\langle\hat t_{2,0,-\sigma}^{\phantom{\dagger}}
 \hat c_{0,\sigma}^{\phantom{\dagger}};
              \hat c_{0,\sigma}^{\dagger}\rangle\rangle
+\,z t
\,\langle\langle\hat t_{1,1^{\prime},-\sigma}^{\phantom{\dagger}}
 \hat c_{0,\sigma}^{\phantom{\dagger}};
              \hat c_{0,\sigma}^{\dagger}\rangle\rangle
 \\  \ &+&\
    2\, t
\,\langle\langle\hat n_{1,-\sigma}^{\phantom{\dagger}}
 \hat c_{0,\sigma}^{\phantom{\dagger}};
              \hat c_{0,\sigma}^{\dagger}\rangle\rangle
-\, 2\, t
\,\langle\langle\hat n_{0,-\sigma}^{\phantom{\dagger}}
 \hat c_{0,\sigma}^{\phantom{\dagger}};
              \hat c_{0,\sigma}^{\dagger}\rangle\rangle
 \\  \ &-&\
     U/2
\,\langle\langle\hat d_{1  , \sigma}^{\phantom{\dagger}}
 \hat t_{1,0,-\sigma}^{\phantom{\dagger}}
 \hat c_{0,\sigma}^{\phantom{\dagger}};
              \hat c_{0,\sigma}^{\dagger}\rangle\rangle,
\label{motion_3_1}
\end{array}
\end{equation}
where we have used various commutator-relations, in particular
Eq.\ (\ref{commutator_5}).
We can simplify Eq.\ (\ref{motion_3_1}) in infinite-dimensions
where the decoupling
\begin{equation}
\,\langle\langle\hat n_{1,-\sigma}^{\phantom{\dagger}}
 \hat c_{0,\sigma}^{\phantom{\dagger}};
                  \hat c_{0,\sigma}^{\dagger}\rangle\rangle
     \ \rightarrow\
<\hat n_{1,-\sigma}^{\phantom{\dagger}}>
\,\langle\langle\hat c_{0,\sigma}^{\phantom{\dagger}};
                   \hat c_{0,\sigma}^{\dagger}\rangle\rangle
\label{d_3}
\end{equation}
becomes exact. Again
$<\hat n_{1,-\sigma}^{\phantom{\dagger}}>\,
  \equiv1/2+\Delta n_{1,-\sigma}$
and using Eq.\ (\ref{definitions}),
$\ 2\hat n_{0,-\sigma}^{\phantom{\dagger}} -1 =
   \hat d_{0,-\sigma}^{\phantom{\dagger}}$,
we can rewrite
\begin{equation}
    2\, t
\,\langle\langle\hat n_{1,-\sigma}^{\phantom{\dagger}}
 \hat c_{0,\sigma}^{\phantom{\dagger}};
                \hat c_{0,\sigma}^{\dagger}\rangle\rangle
-\, 2\, t
\,\langle\langle\hat n_{0,-\sigma}^{\phantom{\dagger}}
 \hat c_{0,\sigma}^{\phantom{\dagger}};
                \hat c_{0,\sigma}^{\dagger}\rangle\rangle
       \ = \
    2\, t\,\Delta n_{1,-\sigma}
\,\langle\langle\hat c_{0,\sigma}^{\phantom{\dagger}};
                \hat c_{0,\sigma}^{\dagger}\rangle\rangle
-\,     t
\,\langle\langle\hat d_{0,-\sigma}^{\phantom{\dagger}}
 \hat c_{0,\sigma}^{\phantom{\dagger}};
                \hat c_{0,\sigma}^{\dagger}\rangle\rangle
\label{e_3}
\end{equation}
in Eq.\ (\ref{motion_3_1}).
In third order we have with the equation of motion for
$\,\langle\langle\hat d_{0,-\sigma}^{\phantom{\dagger}}
 \hat c_{1,\sigma}^{\phantom{\dagger}};
 \hat c_{0,\sigma}^{\dagger}\rangle\rangle$
a second equation of motion,
\begin{equation}
\begin{array}{rcl}
(i\omega_n+\Delta\mu)\,
 \langle\langle\hat d_{0,-\sigma}^{\phantom{\dagger}}
 \hat c_{1,\sigma}^{\phantom{\dagger}};
          \hat c_{0,\sigma}^{\dagger}\rangle\rangle
     \ &=&\  z t
\,\langle\langle\hat d_{0,-\sigma}^{\phantom{\dagger}}
 \hat c_{2,\sigma}^{\phantom{\dagger}};
          \hat c_{0,\sigma}^{\dagger}\rangle\rangle
        -2\, z t
\,\langle\langle\hat j_{1^{\prime},0,-\sigma}^{\phantom{\dagger}}
 \hat c_{1,\sigma}^{\phantom{\dagger}};
          \hat c_{0,\sigma}^{\dagger}\rangle\rangle
  \\ \ &+&\  \phantom{z} t
\,\langle\langle\hat d_{0,-\sigma}^{\phantom{\dagger}}
 \hat c_{0,\sigma}^{\phantom{\dagger}};
          \hat c_{0,\sigma}^{\dagger}\rangle\rangle
      +\, U/2
\,\langle\langle\hat d_{0,-\sigma}^{\phantom{\dagger}}
 \hat d_{1,-\sigma}^{\phantom{\dagger}}
 \hat c_{1,\sigma}^{\phantom{\dagger}};
          \hat c_{0,\sigma}^{\dagger}\rangle\rangle.
\label{motion_3_2}
\end{array}
\end{equation}
A total of six new Greens's functions are generated
in third order by
Eq.\ (\ref{motion_3_1}) and Eq.\ (\ref{motion_3_2}),
$\,\langle\langle\hat d_{0,-\sigma}^{\phantom{\dagger}}
  \hat d_{1,-\sigma}^{\phantom{\dagger}}
  \hat c_{1,\sigma}^{\phantom{\dagger}};
  \hat c_{0,\sigma}^{\dagger}\rangle\rangle$
and
$\,\langle\langle\hat d_{0,-\sigma}^{\phantom{\dagger}}
  \hat c_{2,\sigma}^{\phantom{\dagger}};
  \hat c_{0,\sigma}^{\dagger}\rangle\rangle$,
by Eq.\ (\ref{motion_3_2}) alone,
$\,\langle\langle\hat d_{1,\sigma}^{\phantom{\dagger}}
  \hat t_{1,0,-\sigma}^{\phantom{\dagger}}
  \hat c_{0,\sigma}^{\phantom{\dagger}};
  \hat c_{0,\sigma}^{\dagger}\rangle\rangle$,
$\,\langle\langle\hat t_{2,0,-\sigma}^{\phantom{\dagger}}
  \hat c_{0,\sigma}^{\phantom{\dagger}};
  \hat c_{0,\sigma}^{\dagger}\rangle\rangle$
and
$\,\langle\langle\hat t_{1,1^{\prime},-\sigma}^{\phantom{\dagger}}
  \hat c_{0,\sigma}^{\phantom{\dagger}};
  \hat c_{0,\sigma}^{\dagger}\rangle\rangle$
by Eq.\ (\ref{motion_3_1}) alone, and
$\,\langle\langle\hat j_{1,0,-\sigma}^{\phantom{\dagger}}
  \hat c_{1^{\prime},\sigma}^{\phantom{\dagger}};
  \hat c_{0,\sigma}^{\dagger}\rangle\rangle$
by both Eq.\ (\ref{motion_3_1}) and Eq.\ (\ref{motion_3_2}). We
present the equation of motion for these six Green's functions in
Appendix A. Now we specialize on half-filling, where $\Delta\mu=0$.
For an overview we rewrite here the first two equation of motion,
Eq.\ (\ref{motion_1}) and Eq.\ (\ref{motion_2}):
\begin{eqnarray}
\left[i\omega_n-\tilde t^2 G(i\omega_n)\right]
                           G(i\omega_n)  \ &=&\ 1 + U/2
\,\langle\langle\hat d_{0,-\sigma}^{\phantom{\dagger}}
 \hat c_{0,\sigma}^{\phantom{\dagger}};
 \hat c_{0,\sigma}^{\dagger}\rangle\rangle
\label{half_1}  \\
i\omega_n\,\langle\langle\hat d_{0,-\sigma}^{\phantom{\dagger}}
 \hat c_{0,\sigma}^{\phantom{\dagger}};
 \hat c_{0,\sigma}^{\dagger}\rangle\rangle
                                         \ &=&\
   U/2\, G(i\omega_n) -2\, z t
\,\langle\langle\hat j_{1,0,-\sigma}^{\phantom{\dagger}}
 \hat c_{0,\sigma}^{\phantom{\dagger}};
 \hat c_{0,\sigma}^{\dagger}\rangle\rangle
                       +\, z t
\,\langle\langle\hat d_{0,-\sigma}^{\phantom{\dagger}}
 \hat c_{1,\sigma}^{\phantom{\dagger}};
 \hat c_{0,\sigma}^{\dagger}\rangle\rangle.
\label{half_2}
\end{eqnarray}
Noting that there is no current flowing in equilibrium,
$<\hat c_{1,-\sigma}^{\dagger}
  \hat c_{0,-\sigma}^{\phantom{\dagger}}
 -\hat c_{0,-\sigma}^{\dagger}
  \hat c_{1,-\sigma}^{\phantom{\dagger}}>\,
 \equiv 0$ and
using Eq.\ (\ref{e_3}) with $\Delta n_{1,-\sigma}=0$
the first third-order equation of motion,
Eq.\ (\ref{motion_3_1}), takes the form
\begin{equation}
\begin{array}{rcl}
i\omega_n\,\langle\langle\hat j_{1,0,-\sigma}^{\phantom{\dagger}}
 \hat c_{0,\sigma}^{\phantom{\dagger}};
 \hat c_{0,\sigma}^{\dagger}\rangle\rangle
     \ &=&\ z t
\,\langle\langle\hat j_{1,0,-\sigma}^{\phantom{\dagger}}
 \hat c_{1^{\prime},\sigma}^{\phantom{\dagger}};
 \hat c_{0,\sigma}^{\dagger}\rangle\rangle
        -\, z t
\,\langle\langle\hat t_{2,0,-\sigma}^{\phantom{\dagger}}
 \hat c_{0,\sigma}^{\phantom{\dagger}};
 \hat c_{0,\sigma}^{\dagger}\rangle\rangle
        +\, z t
\,\langle\langle\hat t_{1,1^{\prime},-\sigma}^{\phantom{\dagger}}
 \hat c_{0,\sigma}^{\phantom{\dagger}};
 \hat c_{0,\sigma}^{\dagger}\rangle\rangle
 \\  \ &-&\   t
\,\langle\langle\hat d_{0,-\sigma}^{\phantom{\dagger}}
 \hat c_{0,\sigma}^{\phantom{\dagger}};
 \hat c_{0,\sigma}^{\dagger}\rangle\rangle
    - U/2
\,\langle\langle\hat d_{1  , \sigma}^{\phantom{\dagger}}
 \hat t_{1,0,-\sigma}^{\phantom{\dagger}}
 \hat c_{0,\sigma}^{\phantom{\dagger}};
 \hat c_{0,\sigma}^{\dagger}\rangle\rangle.
\label{half_3_1}
\end{array}
\end{equation}
For completeness, we rewrite also the second third-order
equation of motion,
Eq.\ (\ref{motion_3_1}), for the half-filled case,
\begin{equation}
\begin{array}{rcl}
i\omega_n\,\langle\langle\hat d_{0,-\sigma}^{\phantom{\dagger}}
 \hat c_{1,\sigma}^{\phantom{\dagger}};
 \hat c_{0,\sigma}^{\dagger}\rangle\rangle
     \ &=&\  z t
\,\langle\langle\hat d_{0,-\sigma}^{\phantom{\dagger}}
 \hat c_{2,\sigma}^{\phantom{\dagger}};
 \hat c_{0,\sigma}^{\dagger}\rangle\rangle
        -2\, z t
\,\langle\langle\hat j_{1^{\prime},0,-\sigma}^{\phantom{\dagger}}
 \hat c_{1,\sigma}^{\phantom{\dagger}};
 \hat c_{0,\sigma}^{\dagger}\rangle\rangle
  \\ \ &+&\    t
\,\langle\langle\hat d_{0,-\sigma}^{\phantom{\dagger}}
 \hat c_{0,\sigma}^{\phantom{\dagger}};
 \hat c_{0,\sigma}^{\dagger}\rangle\rangle
      +\, U/2
\,\langle\langle\hat d_{0,-\sigma}^{\phantom{\dagger}}
 \hat d_{1,-\sigma}^{\phantom{\dagger}}
 \hat c_{1,\sigma}^{\phantom{\dagger}};
 \hat c_{0,\sigma}^{\dagger}\rangle\rangle.
\label{half_3_2}
\end{array}
\end{equation}
Particle-hole symmetry at half-filling implies that the real part
of $G(\omega)$ is an odd function of frequency. Consequently
Eq.\ (\ref{half_1}) is a even function of frequency,
Eq.\ (\ref{half_2}) an odd function of frequency  and
Eq.\ (\ref{half_3_1}) and Eq.\ (\ref{half_3_2}) even functions
of frequencies. The absence of a (real) constant term on the
right-hand side of Eq.\ (\ref{half_2}) is therefor forced by
symmetry but the absence of constant terms on the
right-hand side of Eq.\ (\ref{half_3_1}) and Eq.\ (\ref{half_3_2})
is accidental.

%
%%%%%%%%%%%%%%%%%%%%%%%%%%%%%%%%%%%%%%%%%%%%%%%%%%%%%%%
%
%%%%%%%%%%%%%%%%%%%%%%%%%%%%%%%%%%%%%%%%%%%%%%%%%%%%%%%
%

\section*{Approximations}

In order to obtain a solution for $G(i\omega_n)$ we
will decouple the equations of motion. We present here
a simple, systematic decoupling scheme, which may be
applied at any order of the equation of motion. We
generalize the decoupling for
$\,\langle\langle\hat c_{1,\sigma}^{\phantom{\dagger}};
    \hat c_{0,\sigma}^{\dagger}\rangle\rangle$
presented in Eq.\ (\ref{d_1}) by
\begin{equation}
\begin{array}{rcl}
\,\langle\langle\hat j_{1,0,-\sigma}^{\phantom{\dagger}}
 \hat c_{0,\sigma}^{\phantom{\dagger}};
 \hat c_{0,\sigma}^{\dagger}\rangle\rangle
         \ &\rightarrow&\
          -\, t\, G(i\omega_n)
\,\langle\langle\hat d_{0,-\sigma}^{\phantom{\dagger}}
 \hat c_{0,\sigma}^{\phantom{\dagger}};
 \hat c_{0,\sigma}^{\dagger}\rangle\rangle
                \\
\,\langle\langle\hat d_{0,-\sigma}^{\phantom{\dagger}}
 \hat c_{1,\sigma}^{\phantom{\dagger}};
 \hat c_{0,\sigma}^{\dagger}\rangle\rangle
         \ &\rightarrow&\
 \phantom{-}\, t\, G(i\omega_n)
\,\langle\langle\hat d_{0,-\sigma}^{\phantom{\dagger}}
 \hat c_{0,\sigma}^{\phantom{\dagger}};
 \hat c_{0,\sigma}^{\dagger}\rangle\rangle,
\label{d_t}
\end{array}
\end{equation}
which may be motivated by considering Eq.\ (\ref{half_3_1})
and Eq.\ (\ref{half_3_2}). In contrast to
Eq.\ (\ref{d_1}), which is exact in the limit of
infinite dimensions, Eq.\ (\ref{d_t}) is an
approximation only, valid in the paramagnetic state.
In a state with staggered antiferromagnetism the
replacemment
$ G(i\omega_n) \rightarrow G_{-\sigma}(i\omega_n)
$
should be made on the right-hand side of
Eq.\ (\ref{d_t}).
Using Eq.\ (\ref{half_2}) and Eq.\ (\ref{half_1})
we then obtain
\begin{equation}
\left[i\omega_n-\tilde t^2 G(i\omega_n)\right]
                           G(i\omega_n)  \ =\ 1 +
  \frac{(U/2)^2}{i\omega_n-3\,\tilde t^2 G(i\omega_n)}
                           G(i\omega_n).
\label{approx_2}
\end{equation}
Comparing Eq.\ (\ref{approx_2}) with the
Dyson equation of the infinite-dimensional
Bethe-lattice, which has the form
\begin{equation}
\left[i\omega_n+\mu-\Sigma(i\omega_n)-\tilde t^2 G(i\omega_n)\right]
                           G(i\omega_n)  \ =\ 1,
\label{Dyson_equation}
\end{equation}
we find
\begin{equation}
\Sigma^{(2)}(i\omega_n)-\mu \ =\
  \frac{(U/2)^2}{i\omega_n-3\,\tilde t^2 G(i\omega_n)}.
\label{Sigma_2}
\end{equation}
Eq.\ (\ref{Sigma_2})
is the well-know result of the Hubbard-III approximation
\cite{Hubbard_III}.
Here we use the index $^{(2)}$ to indicate that for this
approximation to the self-energy the first equation of
motion has been retained exactly and that the
equation of motion of second order has been decoupled.
In order to obtain $\Sigma^{(3)}(i\omega_n)$
we have to consider the equation of motion of third order,
Eq.\ (\ref{half_3_1}) and Eq.\ (\ref{half_3_2}). For the
decoupling of the terms $\sim t$ on the respective
right-hand sides straightforward generalization of
Eq.\ (\ref{d_t}) can be used. For the terms $\sim U/2$
we proposed the dcoupling scheme
\begin{equation}
\begin{array}{rcl}
\,\langle\langle\hat d_{1,\sigma}^{\phantom{\dagger}}
 \hat t_{1,0,-\sigma}^{\phantom{\dagger}}
 \hat c_{0,\sigma}^{\phantom{\dagger}};
 \hat c_{0,\sigma}^{\dagger}\rangle\rangle
         \ &\rightarrow&\
         - \, (2/U)\,(\Sigma(i\omega_n)-\mu)
\,\langle\langle\hat j_{1,0,-\sigma}^{\phantom{\dagger}}
 \hat c_{0,\sigma}^{\phantom{\dagger}};
 \hat c_{0,\sigma}^{\dagger}\rangle\rangle
                \\
\,\langle\langle\hat d_{0,-\sigma}^{\phantom{\dagger}}
 \hat d_{1,-\sigma}^{\phantom{\dagger}}
 \hat c_{1,\sigma}^{\phantom{\dagger}};
 \hat c_{0,\sigma}^{\dagger}\rangle\rangle
         \ &\rightarrow&\
 \phantom{-}\, (2/U)\,(\Sigma(i\omega_n)-\mu)
\,\langle\langle\hat d_{0,-\sigma}^{\phantom{\dagger}}
 \hat c_{1,\sigma}^{\phantom{\dagger}};
 \hat c_{0,\sigma}^{\dagger}\rangle\rangle,
\label{d_U}
\end{array}
\end{equation}
valid in the half-filled case,
where we have been motivated by Eq.\ (\ref{motion_4_1})
and Eq.\ (\ref{motion_4_2}).
Eq.\ (\ref{half_3_1}) and Eq.\ (\ref{half_3_2}) then become
\begin{equation}
\begin{array}{rcl}
i\omega_n\,\langle\langle\hat j_{1,0,-\sigma}^{\phantom{\dagger}}
 \hat c_{0,\sigma}^{\phantom{\dagger}};
 \hat c_{0,\sigma}^{\dagger}\rangle\rangle
         \ &=&\
   t \,\langle\langle\hat d_{0,-\sigma}^{\phantom{\dagger}}
 \hat c_{0,\sigma}^{\phantom{\dagger}};
 \hat c_{0,\sigma}^{\dagger}\rangle\rangle
    \\   \ &+&\
  3\, \tilde t^2\, G(i\omega_n)
         \,\langle\langle\hat j_{1,0,-\sigma}^{\phantom{\dagger}}
 \hat c_{0,\sigma}^{\phantom{\dagger}};
 \hat c_{0,\sigma}^{\dagger}\rangle\rangle
+                 (\Sigma(i\omega_n)-\mu)
         \,\langle\langle\hat j_{1,0,-\sigma}^{\phantom{\dagger}}
 \hat c_{0,\sigma}^{\phantom{\dagger}};
 \hat c_{0,\sigma}^{\dagger}\rangle\rangle
                  \\
i\omega_n\,\langle\langle\hat d_{0,-\sigma}^{\phantom{\dagger}}
 \hat c_{1,\sigma}^{\phantom{\dagger}};
 \hat c_{0,\sigma}^{\dagger}\rangle\rangle
         \ &=&\
\phantom{-} t
  \,\langle\langle\hat d_{0,-\sigma}^{\phantom{\dagger}}
 \hat c_{0,\sigma}^{\phantom{\dagger}};
 \hat c_{0,\sigma}^{\dagger}\rangle\rangle
    \\   \ &+&\
  3\, \tilde t^2\, G(i\omega_n)
         \,\langle\langle\hat d_{0,-\sigma}^{\phantom{\dagger}}
 \hat c_{1,\sigma}^{\phantom{\dagger}};
 \hat c_{0,\sigma}^{\dagger}\rangle\rangle
+                 (\Sigma(i\omega_n)-\mu)
         \,\langle\langle\hat d_{0,-\sigma}^{\phantom{\dagger}}
 \hat c_{1,\sigma}^{\phantom{\dagger}};
 \hat c_{0,\sigma}^{\dagger}\rangle\rangle.
\label{approx_3_1}
\end{array}
\end{equation}
We replace $\Sigma(i\omega_n)\rightarrow\Sigma^{(3)}(i\omega_n)$
in Eq.\ (\ref{approx_3_1}) and obtain for
   Eq.\ (\ref{half_2})
\[
i\omega_n\,\langle\langle\hat d_{0,-\sigma}^{\phantom{\dagger}}
 \hat c_{0,\sigma}^{\phantom{\dagger}};
 \hat c_{0,\sigma}^{\dagger}\rangle\rangle
  \ = \ U/2\, G(i\omega_n) +
  \frac{3\,\tilde t^2}{i\omega_n-3\,\tilde t^2 G(i\omega_n)
                                -(\Sigma^{(3)}(i\omega_n)-\mu)}
\,\langle\langle\hat d_{0,-\sigma}^{\phantom{\dagger}}
 \hat c_{0,\sigma}^{\phantom{\dagger}};
 \hat c_{0,\sigma}^{\dagger}\rangle\rangle.
\]
Finally, using
Eq.\ (\ref{Dyson_equation}) and Eq.\ (\ref{half_1})
we obtain
\begin{equation}
\Sigma^{(3)}(i\omega_n)-\mu \ = \
 \frac{(U/2)^2}{\displaystyle
i\omega_n
  -\frac{3\,\tilde t^2}{i\omega_n-3\,\tilde t^2 G(i\omega_n)
                                -  (\Sigma^{(3)}(i\omega_n)-\mu)}
               },
\label{Sigma_3}
\end{equation}
with $\mu=U/2$ at half-filling.
In Appendix A we derive the expression for
$\Sigma^{(4)}(i\omega_n)$, see
Eq.\ (\ref{Sigma_4}). Considering the
self-energy expanded in powers of $(U/2)^2$ and
of $\tilde t^2$,
we see that
$\Sigma^{(2)}(i\omega_n)$ is exact up
to $\tilde t^2$ and all powers of $(U/2)^2$.
Similary $\Sigma^{(3)}(i\omega_n)$ is exact up
to $\tilde t^4$ and $\Sigma^{(4)}(i\omega_n)$ up
to $\tilde t^6$.

%
%%%%%%%%%%%%%%%%%%%%%%%%%%%%%%%%%%%%%%%%%%%%%%%%%%%%%%%
%
%%%%%%%%%%%%%%%%%%%%%%%%%%%%%%%%%%%%%%%%%%%%%%%%%%%%%%%
%

\section*{Results}

The solution for the retarded Green's function,
$G(\omega)$, is given by
Eq.\ (\ref{Dyson_equation}) together with the
respective self-energy, see
Eq.\ (\ref{Sigma_2}), Eq.\ (\ref{Sigma_3})
and Eq.\ (\ref{Sigma_4}). For small $U$'s the
solution is metallic, for large $U$'s insulating.
The transition point can be calculated from a small-$\omega$
expansion, by noting that in the insulating state
the Laurent-series of the self-energy starts by
a $1/\omega$ divergence \cite{star_of_stars}:
$\Sigma(\omega)\sim \alpha/\omega + \dots $. From
Dysons"s equation,
Eq.\ (\ref{Dyson_equation}), it follows than that
that
$G(\omega)\sim -\omega/\alpha + \dots $. We find then
that
\begin{equation}
\begin{array}{rcl}
\alpha^{(2)} \  = \
\alpha^{(3)} \ &=&\
\frac{\displaystyle 1}{\displaystyle 4}
      \left(U^2-12\,\tilde t^2 \right) \\
\alpha^{(4)} \ &=&\
\frac{\displaystyle 1}{\displaystyle 4}
\frac{\displaystyle U^2-160\,\tilde t^4/U^2}
     {\displaystyle 1+8\,\tilde t^2/U^2},
\label{alpha_234}
\end{array}
\end{equation}
where $\alpha^{(4)}$ has been calculated in
Appendix A (see Eq.\ (\ref{alpha_4})).
It can be shown \cite{star_of_stars},
considering the spectral representation of
$G(\omega)$, that the coefficient $\alpha\ge0$.
The $U_c$ for the Mott-Hubbard transition is then
given by the $\alpha=0$ condition in
Eq.\ (\ref{alpha_234}). We then find
$U_c^{(2)} = U_c^{(3)} = \sqrt{12}\,\tilde t$ and
$U_c^{(4)}=\sqrt[4]{160}\,\tilde t$. From
now on we use the scaling $\tilde t=1/\sqrt 2$.
We have
\begin{equation}
\begin{array}{rcccl}
U_c^{(2)} = U_c^{(3)} \ &=&\ \sqrt{6} \ &\sim &\ 2.4495 \\
U_c^{(4)} \ &=&\ \sqrt[4]{40} \ &\sim &\ 2.5149.
\label{U_c}
\end{array}
\end{equation}
The $U_c$ given in Eq.\ (\ref{U_c})
has been denoted \cite{IPT_3} $U_{c1}$,
being the critical interation strength at
with the insulating solution becomes stable.
A $U_{c2}$ has also been defined \cite{IPT_3}
as the critical interaction strength at with
the metallic solution becomes unstable, i.e.
when $Im\,G(0)=0$. We will discuss further below
that the equation of motion solutions
always yield
$U_{c2}^{(n)}=U_{c1}^{(n)}\equiv U_c^{(n)}$
as given by Eq.\ (\ref{U_c}), for $n=2,3,4$.

In Fig.\ \ref{alpha_Z}(a) we have plotted
$\alpha^{(3)}$ and $\alpha^{(4)}$ as given by
Eq.\ (\ref{alpha_234}). For comparision we have
included in Fig.\ \ref{alpha_Z} the results obtained
by an exact diagonalization study \cite{star_of_stars},
where the symbol S(1) denotes the so-called Hubbard star
\cite{Hubbard_star} and the symbol S(2) the so-called
star of the stars. In general, the symbol S(n)
denotes clusters which
are truncated Bethe-lattices of order $n$. For
instance, S(1) is the cluster containing a
central site with its $z$ n.n. sites. As the S(n)
are finite clusters, no true Mott-Hubbard transition
is observed, only a crossover. Nevertheless, good
agreement between the S(2) cluster and the
equation of motion results if found for $U\ge3$,
indicating that $U_c\le 3$.

In Fig.\ \ref{alpha_Z}(b) we have plotted the
$Z$-factor, as defined by the seond term in
the Laurent expansion for the self-energy,
$\Sigma(\omega)\sim \alpha/\omega +
 (1-1/Z)\,\omega + \dots $, which we
calculate in Appendix B. In a
Fermi-liquid state ($\alpha=0$) the $Z$-factor
would have the meaning of the inverse effective
mass, $Z=m/m^*$, but in the insulating state
($\alpha>0$)
considered in Fig.\ \ref{alpha_Z}(b)
the $Z$-factor is just a parameter.
$Z$ vanishes at
$U_c$ for all three equation of motion solutions in
the same fashion. To see this we note that
$\alpha^{(n)}\sim U-U_c^{(n)}$
(for $U-U_c^{(n)}\ll U_c^{(n)}$ and $n=2,3,4$, see
Eq.\ (\ref{alpha_234}))
and
$Z^{(n)}\sim [\alpha^{(n)}]^2$ for small
$\alpha^{(n)}$ (see Eq.\ (\ref{Z_23}) and
Eq.\ (\ref{Z_4})). We then obtain
\[
Z^{(n)} \ \sim \ (U-U_c^{(n)})^2
\]
for $n=2,3,4$ and $U_c^{(n)}$ given by
Eq.\ (\ref{U_c}).

In Fig.\ \ref{plot_12} and Fig.\ \ref{plot_34}
we present the equation of
motion results for the density of states,
$-Im\,G^{(n)}(\omega+i\delta)$, for a small
$\delta = 0.0001$. The results for
$U=1$ and $U=2$ are in the metallic state and
for $U=3$ and $U=4$ in the insulating state. The
magnitude of the gap changes little with
$n=1,2,3$, but the shapes of the
Hubbard bands change somewhat
and side bands appear at higher energies.

In Fig.\ \ref{G_Sigma} we present the results for
the density of states at the Fermi level,
$-Im\,G^{(n)}(0)$, and the self-energy at the
Fermi level, $-Im\,\Sigma^{(n)}(0)$,
in the metallic state, as a function of $U$.
Both quantities are related via
$\Sigma^{(n)}(0)=\mu-G^{(n)}(0)/2-1/G^{(n)}(0)$,
compare Eq.\ (\ref{Dyson_equation}), with
$Re\,\Sigma^{(n)}(0)=\mu$. Analytically we find
\begin{equation}
\begin{array}{rcccl}
-Im\,G^{(2)}(0) \ &=&\
   \sqrt{\frac{\displaystyle 12\,\tilde t^2-U^2}
              {\displaystyle     12\,\tilde t^4}
        } \\
-Im\,G^{(3)}(0) \ &=&\
   \sqrt{\frac{\displaystyle 12\,\tilde t^2-U^2}
              {\displaystyle     12\,\tilde t^2
                (\tilde t^2+3/2(U/2)^2)       }
        }.
\label{G(0)_23}
\end{array}
\end{equation}
In a Fermi-liquid state
$Im\,\Sigma(0)\equiv0$ and
$-Im\,G(0)=\sqrt 2$ \cite{Mueller_H},
independent of interaction strength, $U$.
As we see form Fig.\ \ref{G_Sigma}
$-Im\,G^{(n)}(0)<\sqrt 2$ and the
$-Im\,\Sigma^{(n)}(0)>0$. The equation of motion
solutions do not describe a Fermi liquid. Indeed,
in the equation of motion solutions the divergence
\begin{equation}
\lim_{U\to U_c}-Im\,\Sigma^{(n)}(0)\sim
\frac{1}{\sqrt{U_c^{(n)}-U}}
\rightarrow \infty
\label{Sigma_divergence}
\end{equation}
drives the Mott-Hubbard transition. Here
we have derived Eq.\ (\ref{Sigma_divergence}) from
Eq.\ (\ref{G(0)_23}) for $n=2,3$ and verified
it for $n=4$ from a small $G(0)$-expansion of
Eq.\ (\ref{Sigma_4}). We note that the
metallic solution becomes unstable, as seen from
Eq.\ (\ref{Sigma_divergence}), at exactly the same
$U_c^{(n)}$ at which the insulating solution becomes
stable, compare Eq.\ (\ref{U_c}). In an
alternative approach to the Mott-Hubbard
transition on the Bethe lattice in infinite
dimensions \cite{IPT_3} it has been proposed
that the metallic solution might remain stable
up to a much higher $U_{c2}\sim 4.2-4.7$.

%
%%%%%%%%%%%%%%%%%%%%%%%%%%%%%%%%%%%%%%%%%%%%%%%%%%%%%%%
%
%%%%%%%%%%%%%%%%%%%%%%%%%%%%%%%%%%%%%%%%%%%%%%%%%%%%%%%
%

\section*{Conclusions}

We have considered the equation of motion approach
to the Hubbard model on the infinite-dimensional
Bethe-lattice and shown that it is possible to
evaluate higher-order Green's function by a simple
decoupling scheme. We presented analytic and
numerical results for the
half-filled case in the paramagnetice sector.
We found the critical $U_c\sim 2.5$ for
the Mott-Hubbard transition to change only little
with the order of the approximation.  The metallic
state is a non-Fermi-liquid, with a finite
imaginary part of the self-energy at the Fermi-level,
diverging at the Mott-Hubbard transition.
The metallic state does not shown, surprisingly,
any tendency to become more Fermi-liquid like
with increasing order of approximation.

The author would like to thank G. H\"ulsenbeck for
discussions and a carefull reading of the manuscript.
This work was supported by the Minister f\"ur
Wissenschaft und Forschung des Landes
Nordrhein-Westfalen.

%
%%%%%%%%%%%%%%%%%%%%%%%%%%%%%%%%%%%%%%%%%%%%%%%%%%%%%%%
%
%%%%%%%%%%%%%%%%%%%%%%%%%%%%%%%%%%%%%%%%%%%%%%%%%%%%%%%
%

%\begin{verbatim}
\appendix
%\section*{A}
\section{}

Here we present the equations of motion of fourth order
for the six new Green's functions created by the
third-order equations of motion,
Eq.\ (\ref{motion_3_1}) and Eq.\ (\ref{motion_3_2}). The
first one is
\begin{equation}
\begin{array}{rcl}
(i\omega_n+\Delta\mu)
\,\langle\langle\hat d_{0,-\sigma}^{\phantom{\dagger}}
 \hat d_{1,-\sigma}^{\phantom{\dagger}}
 \hat c_{1,\sigma}^{\phantom{\dagger}};
 \hat c_{0,\sigma}^{\dagger}\rangle\rangle
     \ &=&\ U/2
\,\langle\langle\hat d_{0,-\sigma}^{\phantom{\dagger}}
 \hat c_{1,\sigma}^{\phantom{\dagger}};
 \hat c_{0,\sigma}^{\dagger}\rangle\rangle
        +\,2\,   t\,\Delta n_{1,-\sigma}
\,\langle\langle\hat d_{0,-\sigma}^{\phantom{\dagger}}
 \hat c_{0,\sigma}^{\phantom{\dagger}};
 \hat c_{0,\sigma}^{\dagger}\rangle\rangle
 \\  \ &+&\
               z t
\,\langle\langle\hat d_{0,-\sigma}^{\phantom{\dagger}}
 \hat d_{1,-\sigma}^{\phantom{\dagger}}
 \hat c_{2,\sigma}^{\phantom{\dagger}};
 \hat c_{0,\sigma}^{\dagger}\rangle\rangle
        -\,2\, z t
\,\langle\langle\hat j_{1^{\prime},0,-\sigma}^{\phantom{\dagger}}
 \hat d_{1,-\sigma}^{\phantom{\dagger}}
 \hat c_{1,\sigma}^{\phantom{\dagger}};
 \hat c_{0,\sigma}^{\dagger}\rangle\rangle
 \\  \ &-&\
           2\, z t
\,\langle\langle\hat d_{0,-\sigma}^{\phantom{\dagger}}
 \hat j_{2,1,-\sigma}^{\phantom{\dagger}}
 \hat c_{1,\sigma}^{\phantom{\dagger}};
 \hat c_{0,\sigma}^{\dagger}\rangle\rangle,
\label{motion_4_1}
\end{array}
\end{equation}
where we have used the exact decoupling
\[
\,\langle\langle\hat d_{0,-\sigma}^{\phantom{\dagger}}
 \hat d_{1,-\sigma}^{\phantom{\dagger}}
 \hat c_{0,\sigma}^{\phantom{\dagger}};
 \hat c_{0,\sigma}^{\dagger}\rangle\rangle
     \ \rightarrow\
<\hat d_{1,-\sigma}^{\phantom{\dagger}}>
\,\langle\langle\hat d_{0,-\sigma}^{\phantom{\dagger}}
 \hat c_{0,\sigma}^{\phantom{\dagger}};
 \hat c_{0,\sigma}^{\dagger}\rangle\rangle
\]
with
$<\hat d_{1,-\sigma}^{\phantom{\dagger}}>\,=2\Delta n_{1,-\sigma}$.
Note, that the n.n. site $2$ occuring in the last term of the
right-hand side of Eq.\ (\ref{motion_4_1})
in $\hat j_{2,1,-\sigma}^{\phantom{\dagger}}$ could in finite
dimensions also be the central site, $0$, but not in infinite
dimensions. The second equation of motion in fourth order is
\begin{equation}
\begin{array}{rcl}
(i\omega_n+\Delta\mu)
\,\langle\langle\hat d_{1,\sigma}^{\phantom{\dagger}}
 \hat t_{1,0,-\sigma}^{\phantom{\dagger}}
 \hat c_{0,\sigma}^{\phantom{\dagger}};
 \hat c_{0,\sigma}^{\dagger}\rangle\rangle
     \ &=&\
<\hat d_{1,\sigma}^{\phantom{\dagger}}
 \hat t_{1,0,-\sigma}^{\phantom{\dagger}}>
-\,U/2
\,\langle\langle\hat j_{1,0,-\sigma}^{\phantom{\dagger}}
 \hat c_{0,\sigma}^{\phantom{\dagger}};
 \hat c_{0,\sigma}^{\dagger}\rangle\rangle
\\   \ &+&\ z t
\,\langle\langle\hat d_{1,\sigma}^{\phantom{\dagger}}
 \hat t_{1,0,-\sigma}^{\phantom{\dagger}}
 \hat c_{1^{\prime},\sigma}^{\phantom{\dagger}};
                   \hat c_{0,\sigma}^{\dagger}\rangle\rangle
     -\,2\, z t
\,\langle\langle\hat j_{2,1,\sigma}^{\phantom{\dagger}}
 \hat t_{1,0,-\sigma}^{\phantom{\dagger}}
 \hat c_{0,\sigma}^{\phantom{\dagger}};
 \hat c_{0,\sigma}^{\dagger}\rangle\rangle
 \\  \ &-&\
            z t
\,\langle\langle\hat d_{1,\sigma}^{\phantom{\dagger}}
 \hat j_{2,0,-\sigma}^{\phantom{\dagger}}
 \hat c_{0,\sigma}^{\phantom{\dagger}};
 \hat c_{0,\sigma}^{\dagger}\rangle\rangle
        +\, z t
\,\langle\langle\hat d_{1,\sigma}^{\phantom{\dagger}}
 \hat j_{1,1^{\prime},-\sigma}^{\phantom{\dagger}}
 \hat c_{0,\sigma}^{\phantom{\dagger}};
 \hat c_{0,\sigma}^{\dagger}\rangle\rangle,
\label{motion_4_2}
\end{array}
\end{equation}
where we have used, besides others, the commutator relation
$
[\hat t_{1,0,-\sigma}^{\phantom{\dagger}},
 \hat t_{1,0,-\sigma}^{\phantom{\dagger}}] = 0
$.
The third equation of motion in fourth order is
\begin{equation}
\begin{array}{rcl}
(i\omega_n+\Delta\mu)
\,\langle\langle\hat d_{0,-\sigma}^{\phantom{\dagger}}
 \hat c_{2,\sigma}^{\phantom{\dagger}};
 \hat c_{0,\sigma}^{\dagger}\rangle\rangle
     \ &=&\     t
\,\langle\langle\hat d_{0,-\sigma}^{\phantom{\dagger}}
 \hat c_{1,\sigma}^{\phantom{\dagger}};
 \hat c_{0,\sigma}^{\dagger}\rangle\rangle
        +  \, z t
\,\langle\langle\hat d_{0,-\sigma}^{\phantom{\dagger}}
 \hat c_{3,\sigma}^{\phantom{\dagger}};
 \hat c_{0,\sigma}^{\dagger}\rangle\rangle
 \\  \ &-&\
          2\, z t
\,\langle\langle\hat j_{1^{\prime},0,-\sigma}^{\phantom{\dagger}}
 \hat c_{2,\sigma}^{\phantom{\dagger}};
 \hat c_{0,\sigma}^{\dagger}\rangle\rangle
      +\, U/2
\,\langle\langle\hat d_{0,-\sigma}^{\phantom{\dagger}}
 \hat d_{2,-\sigma}^{\phantom{\dagger}}
 \hat c_{2,\sigma}^{\phantom{\dagger}};
 \hat c_{0,\sigma}^{\dagger}\rangle\rangle.
\label{motion_4_3}
\end{array}
\end{equation}
The fourth equation of motion in fourth order is
\begin{equation}
\begin{array}{rcl}
(i\omega_n+\Delta\mu)
\,\langle\langle\hat j_{1^{\prime},0,-\sigma}^{\phantom{\dagger}}
 \hat c_{1,\sigma}^{\phantom{\dagger}};
 \hat c_{0,\sigma}^{\dagger}\rangle\rangle
     \ &=&\     t
\,\langle\langle\hat j_{1^{\prime},0,-\sigma}^{\phantom{\dagger}}
 \hat c_{0,\sigma}^{\phantom{\dagger}};
 \hat c_{0,\sigma}^{\dagger}\rangle\rangle
        +  \, z t
\,\langle\langle\hat j_{1^{\prime},0,-\sigma}^{\phantom{\dagger}}
 \hat c_{2,\sigma}^{\phantom{\dagger}};
                   \hat c_{0,\sigma}^{\dagger}\rangle\rangle
 \\  \ &+&\
          2\,   t \,\Delta n_{1^{\prime},-\sigma}
\,\langle\langle\hat c_{1,\sigma}^{\phantom{\dagger}};
  \hat c_{0,\sigma}^{\dagger}\rangle\rangle
          -\,   t
\,\langle\langle\hat d_{0,-\sigma}^{\phantom{\dagger}}
 \hat c_{1,\sigma}^{\phantom{\dagger}};
 \hat c_{0,\sigma}^{\dagger}\rangle\rangle
 \\  \ &-&\   z t
\,\langle\langle\hat t_{2^{\prime},0,-\sigma}^{\phantom{\dagger}}
 \hat c_{1,\sigma}^{\phantom{\dagger}};
 \hat c_{0,\sigma}^{\dagger}\rangle\rangle
          +\, z t
\,\langle\langle
  \hat t_{1^{\prime},1^{\prime\prime},-\sigma}^{\phantom{\dagger}}
 \hat c_{1,\sigma}^{\phantom{\dagger}};
 \hat c_{0,\sigma}^{\dagger}\rangle\rangle
 \\  \ &-&\
      U/2
\,\langle\langle\hat d_{1^{\prime},\sigma}^{\phantom{\dagger}}
 \hat t_{1^{\prime},0,-\sigma}^{\phantom{\dagger}}
 \hat c_{1,\sigma}^{\phantom{\dagger}};
 \hat c_{0,\sigma}^{\dagger}\rangle\rangle
+\,   U/2
\,\langle\langle\hat d_{0,\sigma}^{\phantom{\dagger}}
 \hat t_{1^{\prime},0,-\sigma}^{\phantom{\dagger}}
 \hat c_{1,\sigma}^{\phantom{\dagger}};
 \hat c_{0,\sigma}^{\dagger}\rangle\rangle,
 \\  \ &+&\
      U/2
\,\langle\langle\hat j_{1^{\prime},0,-\sigma}^{\phantom{\dagger}}
 \hat d_{1,-\sigma}^{\phantom{\dagger}}
 \hat c_{1,\sigma}^{\phantom{\dagger}};
 \hat c_{0,\sigma}^{\dagger}\rangle\rangle,
\label{motion_4_4}
\end{array}
\end{equation}
where we used the exact relation
\[
2\, t
\,\langle\langle(\hat n_{1^{\prime},-\sigma}^{\phantom{\dagger}}
 -\hat n_{0         ,-\sigma}^{\phantom{\dagger}})\,
 \hat c_{1,\sigma}^{\phantom{\dagger}};
 \hat c_{0,\sigma}^{\dagger}\rangle\rangle
\ = \
2\, t\,\Delta n_{1^{\prime},-\sigma}
\,\langle\langle\hat c_{1,\sigma}^{\phantom{\dagger}};
\hat c_{0,\sigma}^{\dagger}\rangle\rangle
-\, t
\,\langle\langle\hat d_{0,-\sigma}^{\phantom{\dagger}}
 \hat c_{1,\sigma}^{\phantom{\dagger}};
 \hat c_{0,\sigma}^{\dagger}\rangle\rangle,
\]
similar to one valid in Eq.\ (\ref{motion_3_1}) and
Eq.\ (\ref{e_3}).
The fifth equation of motion in fourth order is
\begin{equation}
\begin{array}{rcl}
(i\omega_n+\Delta\mu)
\,\langle\langle\hat t_{1,1^{\prime},-\sigma}^{\phantom{\dagger}}
 \hat c_{0,\sigma}^{\phantom{\dagger}};
 \hat c_{0,\sigma}^{\dagger}\rangle\rangle
     \ &=&\
\,<\hat t_{1,1^{\prime},-\sigma}^{\phantom{\dagger}}>
            +\,   t
\,\langle\langle\hat j_{1,0,-\sigma}^{\phantom{\dagger}}
 \hat c_{0,\sigma}^{\phantom{\dagger}};
 \hat c_{0,\sigma}^{\dagger}\rangle\rangle
           -\,    t
\,\langle\langle\hat j_{0,1^{\prime},-\sigma}^{\phantom{\dagger}}
 \hat c_{0,\sigma}^{\phantom{\dagger}};
                   \hat c_{0,\sigma}^{\dagger}\rangle\rangle
 \\  \ &-&\     z t
\,\langle\langle\hat j_{2,1^{\prime},\sigma}^{\phantom{\dagger}}
 \hat c_{0,\sigma}^{\phantom{\dagger}};
 \hat c_{0,\sigma}^{\dagger}\rangle\rangle
            +\, z t
\,\langle\langle\hat j_{1,2^{\prime},-\sigma}^{\phantom{\dagger}}
 \hat c_{0,\sigma}^{\phantom{\dagger}};
 \hat c_{0,\sigma}^{\dagger}\rangle\rangle
 \\  \ &+&\     z t
\,\langle\langle\hat j_{1,1^{\prime},-\sigma}^{\phantom{\dagger}}
 \hat c_{1^{\prime\prime},\sigma}^{\phantom{\dagger}};
 \hat c_{0,\sigma}^{\dagger}\rangle\rangle
+ \, U/2
\,\langle\langle\hat t_{1,1^{\prime},-\sigma}^{\phantom{\dagger}}
 \hat d_{0,-\sigma}^{\phantom{\dagger}};
 \hat c_{0,\sigma}^{\phantom{\dagger}};
 \hat c_{0,\sigma}^{\dagger}\rangle\rangle
 \\  \ &-&\
     U/2
\,\langle\langle\hat d_{1,\sigma}^{\phantom{\dagger}}
 \hat t_{1,1^{\prime},-\sigma}^{\phantom{\dagger}}
 \hat c_{0,\sigma}^{\phantom{\dagger}};
 \hat c_{0,\sigma}^{\dagger}\rangle\rangle
+\,  U/2
\,\langle\langle\hat d_{1^{\prime},\sigma}^{\phantom{\dagger}}
 \hat t_{1,1^{\prime},-\sigma}^{\phantom{\dagger}}
 \hat c_{0,\sigma}^{\phantom{\dagger}};
 \hat c_{0,\sigma}^{\dagger}\rangle\rangle,
\label{motion_4_5}
\end{array}
\end{equation}
which might be further simplified due to
$  t
\,\langle\langle\hat j_{1,0,-\sigma}^{\phantom{\dagger}}
 \hat c_{0,\sigma}^{\phantom{\dagger}};
 \hat c_{0,\sigma}^{\dagger}\rangle\rangle
-\,t
\,\langle\langle\hat j_{0,1^{\prime},-\sigma}^{\phantom{\dagger}}
 \hat c_{0,\sigma}^{\phantom{\dagger}};
                   \hat c_{0,\sigma}^{\dagger}\rangle\rangle \equiv
  2t
\,\langle\langle\hat j_{1,0,-\sigma}^{\phantom{\dagger}}
 \hat c_{0,\sigma}^{\phantom{\dagger}};
 \hat c_{0,\sigma}^{\dagger}\rangle\rangle
$
and
$ - z t
\,\langle\langle\hat j_{2,1^{\prime},\sigma}^{\phantom{\dagger}}
 \hat c_{0,\sigma}^{\phantom{\dagger}};
 \hat c_{0,\sigma}^{\dagger}\rangle\rangle
  + z t
\,\langle\langle\hat j_{1,2^{\prime},-\sigma}^{\phantom{\dagger}}
 \hat c_{0,\sigma}^{\phantom{\dagger}};
 \hat c_{0,\sigma}^{\dagger}\rangle\rangle
\equiv -2zt
\,\langle\langle\hat j_{2,1^{\prime},\sigma}^{\phantom{\dagger}}
 \hat c_{0,\sigma}^{\phantom{\dagger}};
 \hat c_{0,\sigma}^{\dagger}\rangle\rangle
$.
Finally, the sixth equation of motion in fourth order is
\begin{equation}
\begin{array}{rcl}
(i\omega_n+\Delta\mu)
\,\langle\langle\hat t_{2,0,-\sigma}^{\phantom{\dagger}}
 \hat c_{0,\sigma}^{\phantom{\dagger}};
 \hat c_{0,\sigma}^{\dagger}\rangle\rangle
     \ &=&\
\,<\hat t_{2,0,-\sigma}^{\phantom{\dagger}}>
         -\,   t
\,\langle\langle\hat j_{1,0,-\sigma}^{\phantom{\dagger}}
 \hat c_{0,\sigma}^{\phantom{\dagger}};
 \hat c_{0,\sigma}^{\dagger}\rangle\rangle
         +\, z t
\,\langle\langle\hat j_{2,1^{\prime},-\sigma}^{\phantom{\dagger}}
 \hat c_{0,\sigma}^{\phantom{\dagger}};
                   \hat c_{0,\sigma}^{\dagger}\rangle\rangle
 \\  \ &-&\  z t
\,\langle\langle\hat j_{3,0,\sigma}^{\phantom{\dagger}}
 \hat c_{0,\sigma}^{\phantom{\dagger}};
 \hat c_{0,\sigma}^{\dagger}\rangle\rangle
         +\, z t
\,\langle\langle\hat t_{2,0,-\sigma}^{\phantom{\dagger}}
 \hat c_{1^{\prime},\sigma}^{\phantom{\dagger}};
\hat c_{0,\sigma}^{\dagger}\rangle\rangle
 \\  \ &-&\
       U/2
\,\langle\langle\hat d_{2,\sigma}^{\phantom{\dagger}}
 \hat j_{2,0,-\sigma}^{\phantom{\dagger}}
 \hat c_{0,\sigma}^{\phantom{\dagger}};
 \hat c_{0,\sigma}^{\dagger}\rangle\rangle.
\label{motion_4_6}
\end{array}
\end{equation}
Due to particle-hole symmetry the real part of
Eq.\ (\ref{motion_4_2})
is an odd functions of frequency at half filling, where
$\Delta\mu=0$. Therefore the constant term of the
right-hand side of
Eq.\ (\ref{motion_4_2}),
$ <\hat d_{1,\sigma}^{\phantom{\dagger}}
 \hat t_{1,0,-\sigma}^{\phantom{\dagger}}>$,
has to vanish at half-filling, and it does
(compare the definition of
$\hat d_{1,\sigma}^{\phantom{\dagger}}$,
Eq.\ (\ref{definitions})). Note,
that the difference
$ \,\langle\langle\hat t_{1,1^{\prime},-\sigma}^{\phantom{\dagger}}
 \hat c_{0,\sigma}^{\phantom{\dagger}};
 \hat c_{0,\sigma}^{\dagger}\rangle\rangle
-\,\langle\langle\hat t_{2,0,-\sigma}^{\phantom{\dagger}}
 \hat c_{0,\sigma}^{\phantom{\dagger}};
 \hat c_{0,\sigma}^{\dagger}\rangle\rangle
$ is created in the third-order equation of motion,
Eq.\ (\ref{half_3_1}), and that therefore the
difference of the constant terms on the right-hand side of
Eq.\ (\ref{motion_4_5}) and
Eq.\ (\ref{motion_4_6}) respectively, namely
$
 <\hat t_{1,1^{\prime},-\sigma}^{\phantom{\dagger}}>
-<\hat t_{2,0,-\sigma}^{\phantom{\dagger}}>
$
has to vanish at half-filling, and it does.

Now we derive $\Sigma^{(3)}(i\omega_n)$ at half-filling.
Generalizing the decoupling schemes
Eq.\ (\ref{d_t}) and
Eq.\ (\ref{d_U}) we define
\begin{equation}
\begin{array}{rcl}
a(i\omega_n) \ &=& \
\frac{\displaystyle U/2}
     {\displaystyle i\omega_n-5\,\tilde t^2\,G(i\omega_n)
     } \\
b(i\omega_n) \ &=& \
\frac{\displaystyle  t}
     {\displaystyle i\omega_n
     -(\Sigma(i\omega)-\mu)-3\,\tilde t^2\,G(i\omega_n)
     }
\label{a_b}
\end{array}
\end{equation}
and find
\begin{equation}
\begin{array}{rcl}
\,\langle\langle\hat d_{0,-\sigma}^{\phantom{\dagger}}
 \hat d_{1,-\sigma}^{\phantom{\dagger}}
 \hat c_{1,\sigma}^{\phantom{\dagger}};
 \hat c_{0,\sigma}^{\dagger}\rangle\rangle
     \ &=&\
\phantom{-} a(i\omega_n)
\,\langle\langle\hat d_{0,-\sigma}^{\phantom{\dagger}}
 \hat c_{1,\sigma}^{\phantom{\dagger}};
 \hat c_{0,\sigma}^{\dagger}\rangle\rangle
\\
\,\langle\langle\hat d_{1^{\prime},\sigma}^{\phantom{\dagger}}
 \hat t_{1^{\prime},0,-\sigma}^{\phantom{\dagger}}
 \hat c_{0,\sigma}^{\phantom{\dagger}};
 \hat c_{0,\sigma}^{\dagger}\rangle\rangle
     \ &=&\
-a(i\omega_n)
\,\langle\langle\hat j_{1^{\prime},0,-\sigma}^{\phantom{\dagger}}
 \hat c_{0,\sigma}^{\phantom{\dagger}};
 \hat c_{0,\sigma}^{\dagger}\rangle\rangle
\\
\,\langle\langle\hat d_{0,-\sigma}^{\phantom{\dagger}}
 \hat c_{2,\sigma}^{\phantom{\dagger}};
 \hat c_{0,\sigma}^{\dagger}\rangle\rangle
     \ &=&\
\phantom{-}b(i\omega_n)
\,\langle\langle\hat d_{0,-\sigma}^{\phantom{\dagger}}
 \hat c_{1,\sigma}^{\phantom{\dagger}};
 \hat c_{0,\sigma}^{\dagger}\rangle\rangle
\\
\,\langle\langle\hat j_{1^{\prime},0,-\sigma}^{\phantom{\dagger}}
 \hat c_{1,\sigma}^{\phantom{\dagger}};
 \hat c_{0,\sigma}^{\dagger}\rangle\rangle
     \ &=&\
\phantom{-}b(i\omega_n)
\left(
 \,\langle\langle\hat j_{1^{\prime},0,-\sigma}^{\phantom{\dagger}}
  \hat c_{0,\sigma}^{\phantom{\dagger}};
  \hat c_{0,\sigma}^{\dagger}\rangle\rangle
-\,\langle\langle\hat d_{0,-\sigma}^{\phantom{\dagger}}
 \hat c_{1,\sigma}^{\phantom{\dagger}};
 \hat c_{0,\sigma}^{\dagger}\rangle\rangle
\right)
\\
\,\langle\langle
 \hat t_{1^{\prime},1^{\prime\prime},-\sigma}^{\phantom{\dagger}}
 \hat c_{0,\sigma}^{\phantom{\dagger}};
 \hat c_{0,\sigma}^{\dagger}\rangle\rangle
-
\,\langle\langle
 \hat t_{2^{\prime},0,-\sigma}^{\phantom{\dagger}}
 \hat c_{0,\sigma}^{\phantom{\dagger}};
 \hat c_{0,\sigma}^{\dagger}\rangle\rangle
     \ &=&\
2\,b(i\omega_n)
\,\langle\langle\hat j_{1^{\prime},0,-\sigma}^{\phantom{\dagger}}
 \hat c_{0,\sigma}^{\phantom{\dagger}};
 \hat c_{0,\sigma}^{\dagger}\rangle\rangle.
\label{d_4}
\end{array}
\end{equation}
Here we have introduced another decoupling
rule, namely the self-energy $\Sigma(i\omega_n)$ comes in
definitions Eq.\ (\ref{a_b}) only with the prefactor
zero or minus one. We substitute
Eq.\ (\ref{d_4}) into
Eq.\ (\ref{half_3_1}) and
Eq.\ (\ref{half_3_2}). We find
\begin{equation}
\begin{array}{rcl}
  A(i\omega_n)
\,\langle\langle\hat j_{1^{\prime},0,-\sigma}^{\phantom{\dagger}}
 \hat c_{0,\sigma}^{\phantom{\dagger}};
 \hat c_{0,\sigma}^{\dagger}\rangle\rangle
+ B(i\omega_n)
\,\langle\langle\hat d_{0,-\sigma}^{\phantom{\dagger}}
 \hat c_{1,\sigma}^{\phantom{\dagger}};
 \hat c_{0,\sigma}^{\dagger}\rangle\rangle
      \ &=&\ -\, t
\,\langle\langle\hat d_{0,-\sigma}^{\phantom{\dagger}}
 \hat c_{0,\sigma}^{\phantom{\dagger}};
 \hat c_{0,\sigma}^{\dagger}\rangle\rangle
        \\
  2\,B(i\omega_n)
\,\langle\langle\hat j_{1^{\prime},0,-\sigma}^{\phantom{\dagger}}
 \hat c_{0,\sigma}^{\phantom{\dagger}};
 \hat c_{0,\sigma}^{\dagger}\rangle\rangle
+ A(i\omega_n)
\,\langle\langle\hat d_{0,-\sigma}^{\phantom{\dagger}}
 \hat c_{1,\sigma}^{\phantom{\dagger}};
 \hat c_{0,\sigma}^{\dagger}\rangle\rangle
      \ &=&\ \phantom{-}\, t
\,\langle\langle\hat d_{0,-\sigma}^{\phantom{\dagger}}
 \hat c_{0,\sigma}^{\phantom{\dagger}};
 \hat c_{0,\sigma}^{\dagger}\rangle\rangle,
\label{2_by_2}
\end{array}
\end{equation}
with $B(i\omega_n)=z t\,b(i\omega_n)$ and
$A(i\omega_n)\ =\ i\omega_n - U/2\,a(i\omega_n)
- 3\,z t\,b(i\omega_n)$, with
$a(i\omega_n)$ and $b(i\omega_n)$ given by
Eq.\ (\ref{a_b}).
Inverting Eq.\ (\ref{2_by_2}) we obtain
\begin{equation}
\begin{array}{rcl}
\,\langle\langle\hat j_{1^{\prime},0,-\sigma}^{\phantom{\dagger}}
 \hat c_{0,\sigma}^{\phantom{\dagger}};
 \hat c_{0,\sigma}^{\dagger}\rangle\rangle
              \ &=&\ -\, t
\frac{\displaystyle A(i\omega_n)+B(i\omega_n) }
     {\displaystyle A^2(i\omega_n)-2\,B^2(i\omega_n) }
\,\langle\langle\hat d_{0,-\sigma}^{\phantom{\dagger}}
 \hat c_{0,\sigma}^{\phantom{\dagger}};
 \hat c_{0,\sigma}^{\dagger}\rangle\rangle
               \\
\,\langle\langle\hat d_{0,-\sigma}^{\phantom{\dagger}}
 \hat c_{1,\sigma}^{\phantom{\dagger}};
 \hat c_{0,\sigma}^{\dagger}\rangle\rangle
              \ &=&\ \phantom{-}\, t
\frac{\displaystyle A(i\omega_n)+2\,B(i\omega_n) }
     {\displaystyle A^2(i\omega_n)-2\,B^2(i\omega_n) }
\,\langle\langle\hat d_{0,-\sigma}^{\phantom{\dagger}}
 \hat c_{0,\sigma}^{\phantom{\dagger}};
 \hat c_{0,\sigma}^{\dagger}\rangle\rangle.
\label{sol_2}
\end{array}
\end{equation}
Finnally, using Eq.\ (\ref{half_2}), Eq.\ (\ref{half_1}) and
Eq.\ (\ref{Dyson_equation}) we obtain
\begin{equation}
\Sigma^{(4)}(i\omega_n)-\mu
              \ =\
\frac{ (U/2)^2}{i\omega_n -\, \tilde t^2\,
\frac{\displaystyle 2\,A(i\omega_n)+3\,B(i\omega_n) }
     {\displaystyle A^2(i\omega_n)-2\,B^2(i\omega_n) }
               }.
\label{Sigma_4}
\end{equation}
The critical interaction strength, $U_c^{(4)}$, at zero
temperature may by be obtained
from Eq.\ (\ref{Sigma_4}) by considering
the insulating state. For small real frequencies $\omega$
the the self energy starts like
$\Sigma(\omega)=\alpha/\omega + O(\omega)$ and the
Green's function  like
$G(\omega)=-\omega/\alpha + O(\omega^3)$,
which leads to
$B(\omega)\sim -\omega\,\tilde t^2/\alpha$,
$A(\omega)\sim-(\alpha/\omega)(U/2)^2/(\alpha+5\,\tilde t^2)$
and
$\Sigma^{(4)}(\omega)\sim
(U/2)^2/(\omega_n - 2\,\tilde t^2/A(\omega))$.
One then obtains
\begin{equation}
\alpha^{(4)} \ =\ \frac{U^2/4-40\,\tilde t^4/U^2}
                       {1+8\,\tilde t^2/U^2}.
\label{alpha_4}
\end{equation}
Since $\alpha\ge0$
Eq.\ (\ref{alpha_4}) is valid only for
$U>U_c^{(4)}=\sqrt[4]{160}\,\tilde t$. For the usual
scaling $\tilde t=1/\sqrt 2$ the critical
$U_c^{(4)}=\sqrt[4]{40}\sim 2.5149$, which compares
to
$U_c^{(2)}=U_c^{(3)}=\sqrt{6}\sim 2.4495$.
(compare Eq.\ (\ref{U_c})).

%
%%%%%%%%%%%%%%%%%%%%%%%%%%%%%%%%%%%%%%%%%%%%%%%%%%%%%%%
%
%%%%%%%%%%%%%%%%%%%%%%%%%%%%%%%%%%%%%%%%%%%%%%%%%%%%%%%
%

%\begin{verbatim}
%\appendix
\section{}

Here we consider the insulating state, in
which the Laurent expansion of the retarded
self-energy starts like \cite{star_of_stars}
\begin{equation}
\begin{array}{rcl}
\Sigma(\omega) \ &=&\
\frac{\displaystyle\alpha}{\displaystyle\omega}
               \ +\ \mu \
               \ +\ (1-
\frac{\displaystyle 1}{\displaystyle Z})\,\omega\ +\ \dots \\
G(\omega) \ &=&\ -\
\frac{\displaystyle \omega}{\displaystyle\alpha}
\ - \ (\frac{\displaystyle 1}{\displaystyle\alpha})^2
(\frac{\displaystyle 1}{\displaystyle Z}+
\frac{\displaystyle\tilde t^2}{\displaystyle\alpha})\,
                            \omega^3\ + \ \dots,
\label{Laurent}
\end{array}
\end{equation}
where we have used Eq.\ (\ref{Dyson_equation}) for
self-consistency. In order to calculate $Z$ we need
\begin{equation}
\begin{array}{rcl}
\frac{\displaystyle 1}{\displaystyle\Sigma(\omega)-\mu}
         \  &=&\  \omega/\alpha
\ - \ (1-1/Z)\, \omega^3/\alpha^2
                            \ + \ \dots \\
\frac{\displaystyle 1}{\displaystyle\omega+\mu-\Sigma(\omega)
-3\,\tilde t^2\,G(\omega)}
         \  &=&\  \ - \omega/\alpha
\ - \ (1/Z+3\,\tilde t^2/\alpha)\,
\omega^3/\alpha^2 \ + \ \dots \\
\frac{\displaystyle 1}{\displaystyle\omega
-5\,\tilde t^2\,G(\omega)}
         \  &=&\  \
\frac{\displaystyle\alpha}{\displaystyle(\alpha+5\,\tilde t^2)\omega}
\ - \ 5\,\tilde t^2\,
      \frac{\displaystyle 1/Z+\tilde t^2/\alpha}
           {\displaystyle(\alpha+5\,\tilde t^2)^2}
\, \omega + \ \dots \ .
\label{help_Z}
\end{array}
\end{equation}
Using Eq.\ (\ref{Sigma_2}), Eq.\ (\ref{Sigma_3})
together with
Eq.\ (\ref{Dyson_equation}) and Eq.\ (\ref{help_Z})
one finds
\begin{equation}
\begin{array}{rcl}
Z^{(2)} \ &=&\
[\alpha^{(2)}]^2 \, /\, [(U/2)^2\alpha^{(2)}+3\,\tilde t^4] \\
Z^{(3)} \ &=&\
[\alpha^{(3)}]^2 \, /\, [(U/2)^2\alpha^{(3)}+9\,\tilde t^4].
\label{Z_23}
\end{array}
\end{equation}
For the evaluation of $Z^{(4)}$ we need
\[
\begin{array}{rcl}
\frac{\displaystyle 2A(\omega)+3B(\omega)}
     {\displaystyle A^2(\omega)-2B^2(\omega)}
        \ &=&\
-2\frac{\displaystyle \alpha+5\,\tilde t^2}
     {\displaystyle \alpha (U/2)^2}
 \, \omega
-10\frac{\displaystyle \tilde t^2}
     {\displaystyle \alpha^2 (U/2)^2}
    (1/Z+\,\tilde t^2/\alpha)
 \, \omega^3  \\
        \ &-&\
  \frac{\displaystyle (\alpha+5\,\tilde t^2)^2}
     {\displaystyle \alpha^2 (U/2)^4}
    (2+9\,\tilde t^2/\alpha)
 \, \omega^3 \, + \, \dots \ .
\end{array}
\]
Finnally obtain
\begin{equation}
  \frac{\displaystyle \alpha^{(4)} }
       {\displaystyle Z^{(4)} }
 \left[ 1 + 8\,\tilde t^2/U^2 \right]
   \ =\
 (U/2)^2
+ \,\tilde t^2
  \frac{\displaystyle (\alpha^{(4)}+5\,\tilde t^2)^2}
       {\displaystyle (U/2)^4}
  \left[ 2 + 9\,\tilde t^2/\alpha^{(4)} \right]
+
  \frac{\displaystyle 10\,\tilde t^2}
       {\displaystyle (U/2)^2\alpha^{(4)}},
\label{Z_4}
\end{equation}
from which we may obtain $Z^{(4)}$, whith
Eq.\ (\ref{alpha_4}) for $\alpha^{(4)}$.

%
%%%%%%%%%%%%%%%%%%%%%%%%%%%%%%%%%%%%%%%%%%%%%%%%%%%%%%%
%
%%%%%%%%%%%%%%%%%%%%%%%%%%%%%%%%%%%%%%%%%%%%%%%%%%%%%%%
%

%
%%%%%%%%%%%%%%%%%%%%%%%%%%%%%%%%%%%%%%%%%%%%%%%%%%%%%%%
%
%%%%%%%%%%%%%%%%%%%%%%%%%%%%%%%%%%%%%%%%%%%%%%%%%%%%%%%
%

%
%%%%%%%%%%%%%%%%%%%%%%%%%%%%%%%%%%%%%%%%%%%%%%%%%%%%%%%
%
%%%%%%%%%%%%%%%%%%%%%%%%%%%%%%%%%%%%%%%%%%%%%%%%%%%%%%%
%                                 ^
\begin{figure}
\caption{(a) The coeficient $\alpha$  and (b) the
         coefficient $Z$ of the self-energy,
          $\Sigma(\omega)\sim\alpha/\omega+U/2
           +(1-1/Z)\omega+\dots$, in the
          insulating state at half-filling,
          as a function of $U$.
          Shown are the results for
          $\Sigma^{(2)}(\omega)$ (dotted line),
          $\Sigma^{(3)}(\omega)$ (dashed line) and
          $\Sigma^{(4)}(\omega)$ (solid line).
          For comparision, the corresponding results obtaind
          by exactly diagonalizing the Hubbard star, S(1)
          and the star of the stars, S(2), are given.
          }
\label{alpha_Z}
\end{figure}
\begin{figure}
\caption{ The density of states,
          $-Im\,G^{(2)}(\omega)$ (dotted line),
          $-Im\,G^{(3)}(\omega)$ (dashed line) and
          $-Im\,G^{(4)}(\omega)$ (solid line)
          in the metallic state at half-filling for
          (a) $U=1$ and (b) $U=2$.
          }
\label{plot_12}
\end{figure}
\begin{figure}
\caption{ The density of states,
          $-Im\,G^{(2)}(\omega)$ (dotted line),
          $-Im\,G^{(3)}(\omega)$ (dashed line) and
          $-Im\,G^{(4)}(\omega)$ (solid line)
          in the insulating state at half-filling for
          (a) $U=3$ and (b) $U=4$.
          }
\label{plot_34}
\end{figure}
\begin{figure}
\caption{(a) The density of states at the Fermi level
          in the half-filled metallic state,
          $-Im\,G^{(2)}(0)$ (dotted line),
          $-Im\,G^{(3)}(0)$ (dashed line) and
          $-Im\,G^{(4)}(0)$ (solid line) as a
          function of $U$ and (b)
          the self-energy at the Fermi level,
          $-Im\,\Sigma^{(2)}(0)$ (dotted line),
          $-Im\,\Sigma^{(3)}(0)$ (dashed line) and
          $-Im\,\Sigma^{(4)}(0)$ (solid line).
          }
\label{G_Sigma}
\end{figure}
\end{document}